\documentclass[journal]{IEEEtran}
\IEEEoverridecommandlockouts
\usepackage{cite}
\usepackage{amsmath,amssymb,amsfonts}
\usepackage{algorithmic}
\usepackage{graphicx}
\usepackage{textcomp}
\usepackage{xcolor}
\usepackage{url}
\usepackage{hyperref}
\usepackage{booktabs}
\usepackage{subcaption}
\def\BibTeX{{\rm B\kern-.05em{\sc i\kern-.025em b}\kern-.08em
    T\kern-.1667em\lower.7ex\hbox{E}\kern-.125emX}}
\begin{document}

\title{4D Imaging in ISAC Systems: A Framework Based on 5G NR Downlink Signals}

\author{Haoyang~Weng,
        Haisu~Wu,
        Hong~Ren,~\IEEEmembership{Member,~IEEE,}
        ~Cunhua~Pan,~\IEEEmembership{Senior~Member,~IEEE}
        and~Jiangzhou~Wang,~\IEEEmembership{Fellow,~IEEE}
\thanks{Haoyang Weng, Haisu~Wu, Hong Ren, Cunhua Pan, and Jiangzhou Wang are with National Mobile Communications Research Laboratory, Southeast University, Nanjing, China. (e-mail: 220241245, wuhaisu, hren, cpan, j.z.wang@seu.edu.cn).}
}

\maketitle

\begin{abstract}
Integrated sensing and communication (ISAC) has emerged as a key enabler for sixth-generation (6G) wireless networks, supporting spectrum sharing and hardware integration. Beyond communication enhancement, ISAC also enables high-accuracy environment reconstruction and imaging, which are crucial for applications such as autonomous driving and digital twins. This paper proposes a 4D imaging framework fully compliant with the 5G New Radio (NR) protocol, ensuring compatibility with cellular systems. Specifically, we develop an end-to-end processing chain that covers waveform generation, echo processing, and multi-BS point cloud fusion. Furthermore, we introduce Zoom-OMP, a coarse-to-fine sparse recovery algorithm for high-resolution angle estimation that achieves high accuracy with reduced computational cost. The simulation results demonstrate that the proposed framework achieves robust 4D imaging performance with superior spatial accuracy and reconstruction quality compared to conventional benchmarks, paving the way for practical ISAC-enabled environment reconstruction in 6G networks.
\end{abstract}


\section{Introduction}
The sixth-generation (6G) mobile communication systems are envisioned to empower a series of emerging applications through capabilities such as full-spectrum resources, ultra-large bandwidth, and massive antenna arrays. Among these capabilities, Integrated Sensing and Communication (ISAC) has been recognized as a key enabling technology. By sharing spectrum resources and hardware infrastructures, ISAC not only improves spectral efficiency and reduces deployment costs\cite{liu2022integrated}, but also supports new service scenarios, including vehicle-to-thing (V2X), extended reality (XR), and unmanned aerial vehicle (UAV) management in the low-altitude economy \cite{tang2025cooperative,zhang2024target}. Beyond simply transmitting data, ISAC transforms the communication network into a ubiquitous sensing platform. This paradigm shift allows the realization of high-resolution 4D imaging and environment reconstruction for urban traffic roads, which is significant for applications such as autonomous driving and building the digital world \cite{tan2021integrated}.

In recent years, ISAC-based environment reconstruction and imaging have attracted significant research interest. Existing studies have focused mainly on signal processing and estimation algorithms for extracting channel parameters from orthogonal frequency-division multiplexing (OFDM) signals. For example, early work demonstrated the feasibility of high-precision 3D imaging by analyzing the phase offset of subcarriers in reflected millimeter wave signals to accurately estimate the time-of-flight (ToF) \cite{guan20213}. To further enhance image resolution, subsequent studies have proposed hybrid frameworks that combine traditional 2D-FFT with super-resolution algorithms such as 2D-MUSIC \cite{lu2023isac}. At the system level, architectures based on multiple transmission and reception points (TRPs) have been investigated to capture scattering information and assist in scene reconstruction \cite{zhou20236g}. Moreover, the influence of key ISAC parameters, such as bandwidth and subcarrier spacing, on the quality of the reconstruction has been systematically analyzed \cite{song20243d}. Despite these advances, two critical limitations remain. First, most existing studies rely on idealized OFDM signal models, neglecting the specific physical layer structure of the 5G New Radio (NR) standard. This mismatch hinders their practical deployment in real-world cellular networks. Second, the trade-off between imaging resolution and computational complexity remains a bottleneck for achieving real-time application.

To address these challenges, this paper proposes a 4D imaging framework that is fully compliant with the 5G NR standard. The main contributions of this paper are summarized as follows:
\begin{enumerate}
    \item[1)] First, we design an end-to-end 4D imaging framework that leverages standard 5G NR downlink signals, encompassing the complete processing chain from waveform generation to multi-view point cloud fusion.
    \item[2)] Second, we propose an efficient angle estimation algorithm named Zoom-OMP, which adopts a coarse-to-fine sparse recovery strategy to achieve high angular resolution while reducing computational complexity.

\end{enumerate}

\section{5G NR BASIS}

The foundation of our framework is the use of standard 5G NR downlink signals as sensing sources.  In this section, we briefly introduce the frame structure and physical resources defined in the 3GPP TS 38.211 standard\cite{3gpp_ts38211_v17_5_0}, which serve as essential prerequisites for the subsequent analysis.
\subsection{Frame Structure}
\begin{figure}[htbp]
    \centering
    \includegraphics[width=0.9\columnwidth]{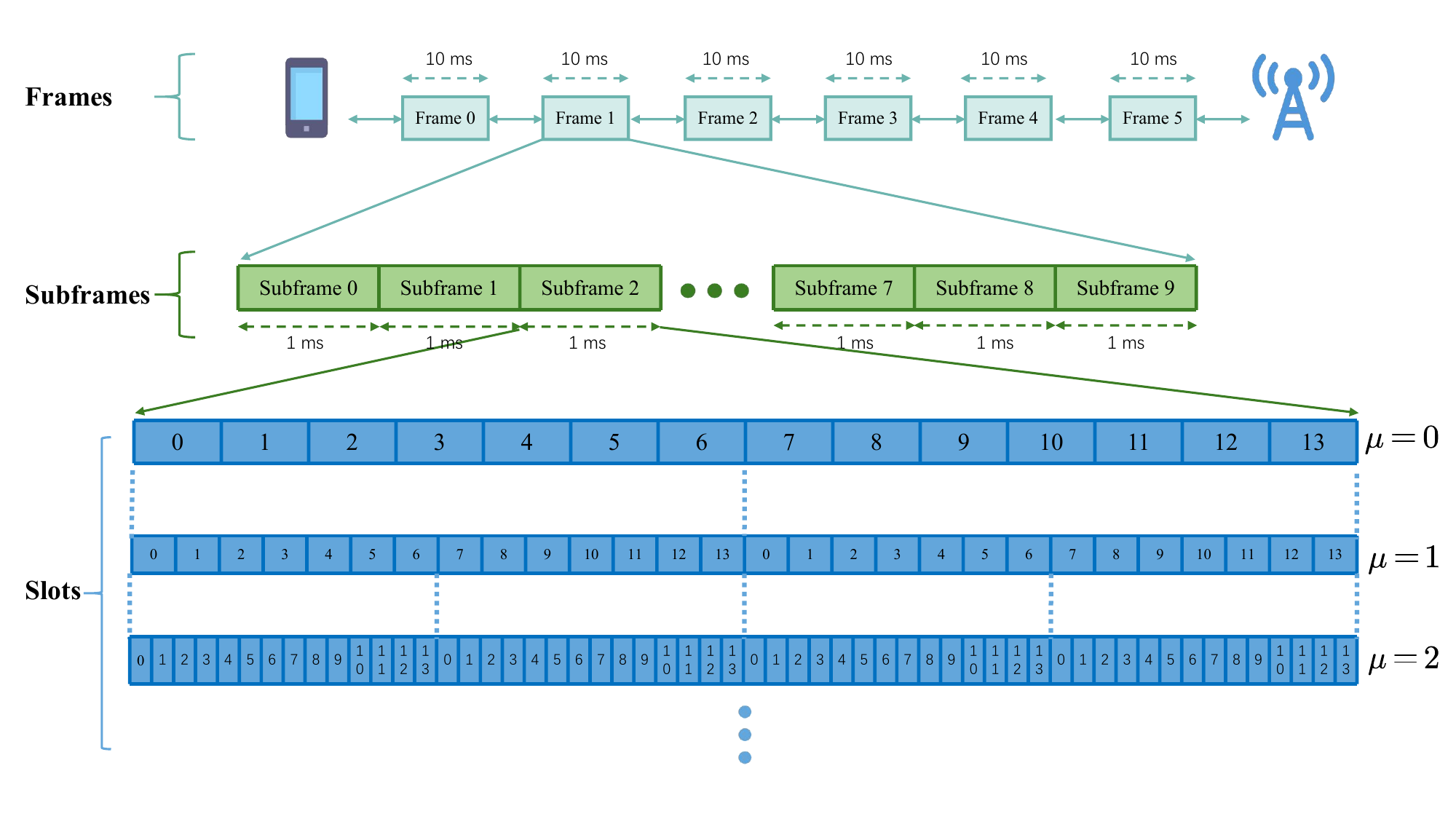}
    \caption{Time-domain structure in 5G NR}
    \label{fig:Frame_Structure}
\end{figure}

In the time domain, 5G NR transmissions are organized into 10 ms radio frames, and each radio frame contains 10 subframes with a duration of 1 ms. A subframe is further divided into multiple slots, the number of which depends on the selected numerology $\mu $. The flexible numerology design allows 5G NR to support different subcarrier spacings (SCS), defined as $\varDelta f^{\mu}=2^{\mu}\times 15 \left[ \mathrm{KHz} \right] $, where $\mu $ can range from 0 to 6. Consequently, a subframe contains $2^{\mu}$ slots. However, regardless of the SCS, each slot consistently consists of 14 (for normal CP) or 12 (for extended CP) OFDM symbols. Fig.~\ref{fig:Frame_Structure} illustrates this hierarchical time-domain structure.
\subsection{Physical Resource}
The physical resources of NR are represented on a two-dimensional time-frequency plane, called the resource grid. In the frequency domain, the basic unit for resource allocation is the resource block (RB), which consists of 12 consecutive subcarriers. In the time domain, the basic unit is a single OFDM symbol. The smallest unit in the resource grid is the resource element (RE), which represents one subcarrier on an OFDM symbol. Due to variations in subcarrier spacing, different resource grids occupy different bandwidths in the frequency domain. NR supports deployment in the sub-6 GHz bands, defined as Frequency Range 1 (FR1), and in the millimeter-wave bands, defined as Frequency Range 2 (FR2). FR1 supports $\mu \in \left\{ 0,1,2 \right\} $ with a maximum bandwidth of 100 MHz \cite{3gpp_ts38101_1_v17_10_0}, while FR2 supports $\mu \in \left\{ 2,3,4 \right\} $ with a maximum bandwidth of up to 400 MHz \cite{3gpp_ts38101_2_v17_11_0}.
\subsection{Physical Channels and Signals}
\begin{figure}[htbp]
    \centering
    \includegraphics[width=0.9\columnwidth]{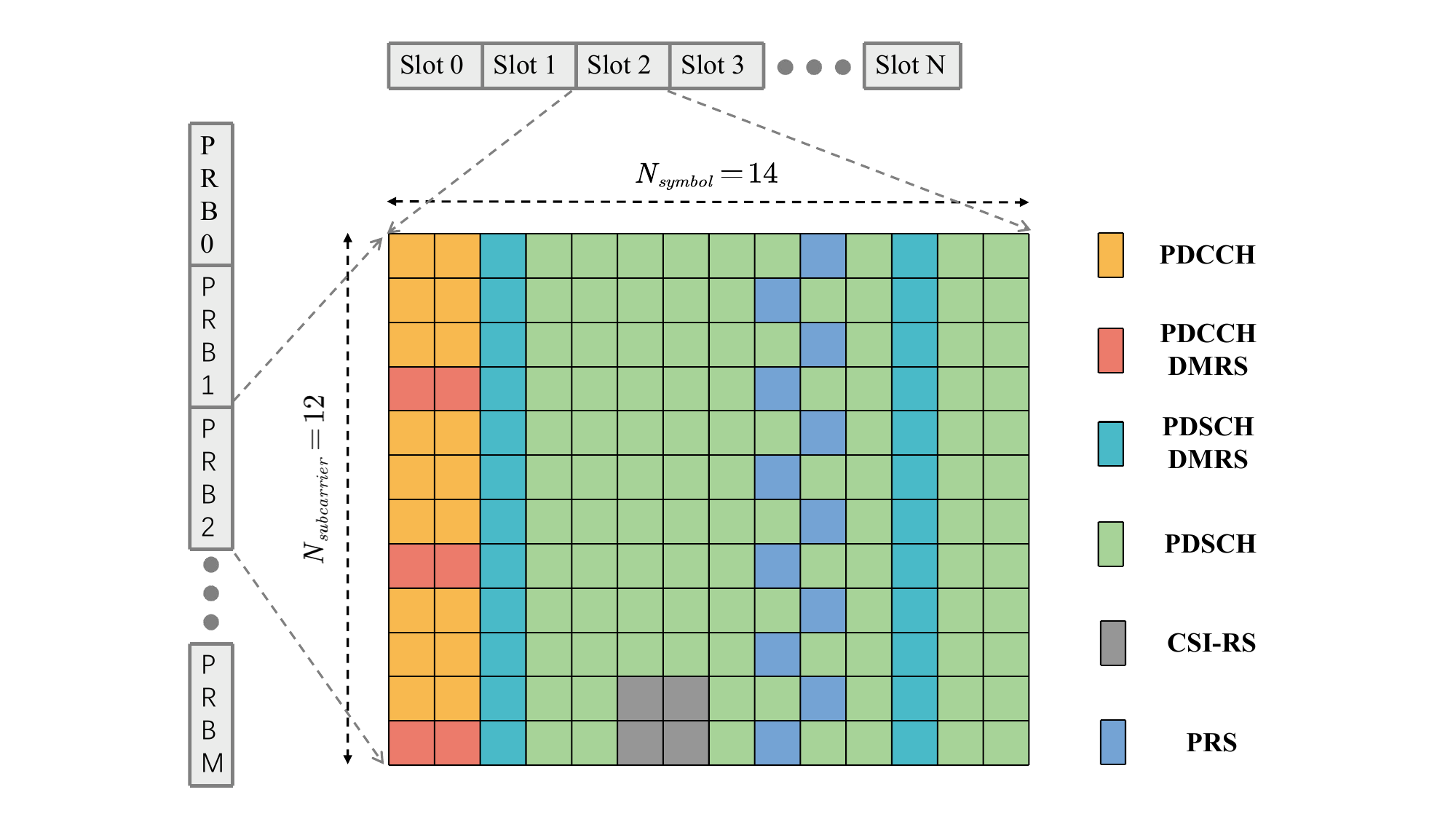}
    \caption{Time-frequency grid within one slot and one PRB}
    \label{fig:Resource_Grid}
\end{figure}

The NR time-frequency resource grid not only carries user data, but also multiplexes various physical channels and reference signals. Fig.~\ref{fig:Resource_Grid}  shows a typical multiplexing scheme of these channels and signals within one slot. The Physical Downlink Control Channel (PDCCH) delivers scheduling information, while the Physical Downlink Shared Channel (PDSCH) carries user data. To support coherent demodulation and channel estimation, NR defines several types of reference signal. In sensing applications, reference signals such as the Demodulation Reference Signal (DMRS), Channel State Information Reference Signal (CSI-RS), and Positioning Reference Signal (PRS) are of particular importance. Their flexible time-frequency configurations form the basis for extracting range, velocity, and angle information from echo signals.

\section{System Model}

\subsection{ISAC Imaging System}

\begin{figure}[htbp]
    \centering
    \includegraphics[width=0.9\columnwidth]{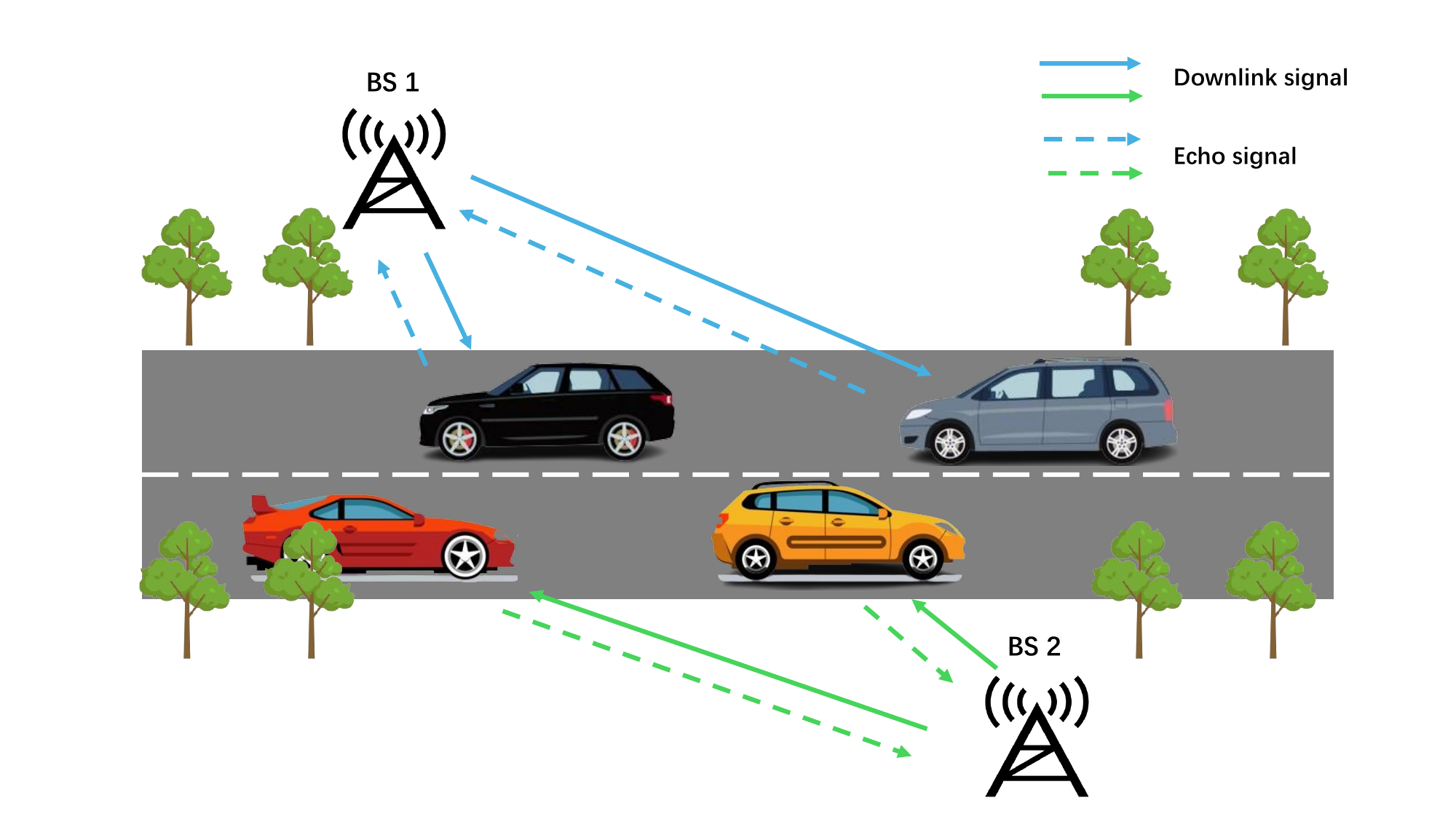}
    \caption{System model for ISAC-based 4D imaging}
    \label{fig:system_model}
\end{figure}

The ISAC-based 4D imaging system model is shown in Fig.~\ref{fig:system_model}. We consider an urban traffic scenario with multiple 5G base stations (BSs) deployed. Each BS is equipped with a pair of uniform planar arrays (UPAs) for transmission and reception. The transmit array generates standard downlink signals that not only serve communication users but also illuminate the surrounding environment. The receive array then captures the echoes reflected from targets such as vehicles and pedestrians, allowing estimation of the key parameters of environmental scatterers, including their range, velocity, and angle. In this cooperative sensing architecture, the spatial diversity provided by multiple BSs enhances detection coverage and mitigates the occlusion problem inherent in single-viewpoint sensing. 

\subsection{Signal Transmission Model}

\textit{1) Antenna Array Model:} Each BS is equipped with a pair of $P\times Q$ UPA placed in the X--Y plane, with $P$ elements along the X-axis and $Q$ elements along the Y-axis. For a plane wave with azimuth $\theta$ and elevation $\varphi$, the steering vector $\mathbf{a}( \theta ,\varphi) \in \mathbb{C} ^{PQ\times 1}$ is expressed as the Kronecker product of the horizontal component $\mathbf{a}_p( \theta ,\varphi) \in \mathbb{C} ^{P\times 1}$ and the vertical component $\mathbf{a}_q( \theta ,\varphi) \in \mathbb{C} ^{Q\times 1}$:
\begin{equation}
\mathbf{a}(\theta, \varphi) = \mathbf{a}_q(\theta, \varphi) \otimes \mathbf{a}_p(\theta, \varphi),
\end{equation}
where $\mathbf{a}_p(\theta, \varphi) = [1, \dots, e^{-j\frac{2\pi d}{\lambda}(P-1)\cos(\varphi)\cos(\theta)}]^T$ and $\mathbf{a}_q(\theta, \varphi) = [1, \dots, e^{-j\frac{2\pi d}{\lambda}(Q-1)\cos(\varphi)\sin(\theta)}]^T$. Here, $d$ denotes the antenna spacing and $\lambda$ denotes the wavelength. The transmit and receive arrays are assumed to have identical structure; thus their steering vectors share the same mathematical form and are denoted as $\boldsymbol{a}_{\mathrm{Tx}}\left( \theta ,\varphi \right) $ and $\boldsymbol{a}_{\mathrm{Rx}}\left( \theta ,\varphi \right) $,  respectively.

\textit{2) Transmitted Signal Model:} In this system, the BS transmits standard 5G NR downlink OFDM signals. In the time-frequency grid, the transmitted single-stream data is represented by a complex matrix $\mathbf{S}\in \mathbb{C} ^{K\times L}$, where $K$ and $L$ denote the number of subcarriers and OFDM symbols, respectively. Each element $s_{k,l}$ is the data symbol allocated to the $k$-th subcarrier and the $l$-th OFDM symbol. Then it is weighted by a digital precoding vector $\boldsymbol{w}\in \mathbb{C} ^{PQ\times 1}$ to generate the transmitted signal vector $\boldsymbol{x}_{k,l}\in \mathbb{C} ^{PQ\times 1}$:
\begin{equation}
\boldsymbol{x}_{k,l}=\boldsymbol{w}s_{k,l}.
\end{equation}

\textit{3) Sensing Channel Model:} The environment is modeled as a collection of $N_s$ discrete scatterers. The signal emitted by the transmit array reflects off the scatterers and is captured by the receive array. The channel matrix in the frequency domain $\mathbf{H}_{k,l}\in \mathbb{C} ^{PQ\times PQ}$ is expressed as
\begin{equation}
\mathbf{H}_{k,l}=\sum_{i=1}^{N_s}{\alpha _i\boldsymbol{a}_{\mathrm{Rx}}\left( \theta _i,\varphi _i \right) \boldsymbol{a}_{\mathrm{Tx}}\left( \theta _i,\varphi _i \right) ^H\cdot}e^{-j2\pi k\varDelta f\tau _i}\cdot e^{j2\pi f_{D,i}lT_s},
\end{equation}
where $\alpha _i$ is the complex channel gain, $\varDelta f$ is the subcarrier spacing, $T_{s}$ is the duration of the OFDM symbol (including the cyclic prefix, CP),  $\tau _i=\frac{2R_i}{c}$ is the propagation delay with $R_i$ denoting the distance between the BS and the $i$-th scatterer, $c$ is the speed of light, and $f_{D,i}=\frac{2v_i}{\lambda}$ is the Doppler shift with $v_i$ denoting the radial velocity of the $i$-th scatterer. 

\textit{4) Received Signal Model: }The received signal vector $\boldsymbol{y}_{k,l}\in \mathbb{C} ^{PQ\times 1}$ is the result of the transmitted signal vector $\boldsymbol{x}_{k,l}$ passing through the channel $\mathbf{H}_{k,l}$ with additive noise: 
\begin{equation}
\boldsymbol{y}_{k,l}=\mathbf{H}_{k,l}\boldsymbol{x}_{k,l}+\boldsymbol{n}_{k,l},
\label{received_signal}
\end{equation}
where $\boldsymbol{n}_{k,l}\in \mathbb{C} ^{PQ\times 1}$ represents the additive white Gaussian noise vector.

\section{ISAC IMAGING SCHEME}
Based on the above signal model, we now describe the framework for environment reconstruction. The process begins with a preliminary estimate of the range and velocity. This is followed by a high-resolution angle estimate. Finally, the imaging results from multiple BSs are fused to generate a global point cloud.

\subsection{Joint Range–Velocity Estimation}
This subsection aims to extract the range and velocity parameters of the scatterers from the echo signals. The signal received at the $(p,q)$-th antenna can be represented
as a 2D signal matrix $\mathbf{Y}_{pq}\in \mathbb{C} ^{K\times L}$, where each element is the corresponding component of the received vector $\boldsymbol{y}_{k,l}$ in \ref{received_signal}, that is, $\left[ \mathbf{Y}_{pq} \right] _{k,l}\triangleq \left[ \boldsymbol{y}_{k,l} \right] _{pq}$, where $\left[ \cdot \right] _{k,l}$ denotes the element at the $k$-th subcarrier and the $l$-th OFDM symbol and $\left[ \cdot \right] _{pq}$ denotes the component corresponding to the $(p,q)$-th antenna. To isolate the channel response, the influence of the known transmitted symbols must be removed. The estimated channel matrix $\hat{\mathbf{H}}_{pq}\in \mathbb{C} ^{K\times L}$ is then obtained by element-wise division, which can be expressed as
\begin{equation}
\left[ \hat{\mathbf{H}}_{pq} \right] _{k,l}=\frac{\left[ \mathbf{Y}_{pq} \right] _{k,l}}{s_{k,l}}.
\end{equation}
We assume that the transmitted symbols are known, thus allowing for element-wise division across all non-zero resource elements. 

To obtain a smoother spectrum for peak detection, zero-padding is applied to the estimated channel matrix. This extends $\hat{\mathbf{H}}_{pq}$ to a larger matrix $\tilde{\mathbf{H}}_{pq}\in \mathbb{C} ^{N_R\times N_D}$, with $N_R\geqslant K$ and $N_D\geqslant L$. We then transform this extended matrix from the time-frequency domain to the delay-Doppler domain to obtain the Range-Doppler Map (RDM). This transformation consists of an Inverse Fast Fourier Transform (IFFT) along the subcarrier dimension and a Fast Fourier Transform (FFT) along the symbol dimension. The RDM element for that antenna $\mathbf{P}_{pq}\in \mathbb{C} ^{N_R\times N_D}$ at $(m,n)$ is calculated as:
\begin{equation}
\left[ \mathbf{P}_{pq} \right] _{m,n}=\sum_{k=0}^{K-1}{\sum_{l=0}^{L-1}{\left[ \tilde{\mathbf{H}}_{pq} \right] _{k,l}\cdot e^{j2\pi \frac{mk}{N_R}}\cdot e^{-j2\pi \frac{nl}{N_D}}}},
\end{equation}
where $m$ and $n$ are the indices for the range and Doppler dimensions, respectively.

To improve the Signal-to-Noise Ratio (SNR), we coherently integrate the RDMs of all $PQ$ antennas to obtain the final RDM $\mathbf{P}$:
\begin{equation}
\mathbf{P}=\left| \sum_{p=0}^{P-1}{\sum_{q=0}^{Q-1}{\mathbf{P}_{pq}}} \right|^2.
\end{equation}
The peaks $( m_i,n_i)$ on the RDM indicate the presence of the $i$-th scatterer. To detect these peaks against the background noise, we adopt the Ordered Statistics Constant False Alarm Rate (OSCA-CFAR) detector\cite{lu2023isac}. For the $i$-th scatterer detected by CFAR, its range $R_i$ and radial velocity $v_i$ are given by:
\begin{equation}
R_i=\frac{cm_i}{2N_R\varDelta f},
\end{equation}
\begin{equation}
v_i=\frac{\lambda n_i}{2N_DT_{s}}.
\end{equation}

\subsection{High-Resolution Angle Estimation}
After obtaining the range and velocity of the scatterers, this section aims to accurately estimate the azimuth angle $\theta$ and the elevation angle $\varphi$ for each scatterer. The RDM values at the peak $\left( m_i,n_i \right) $ across all antennas form the spatial channel vector $\boldsymbol{h}_{spa}^{\left( i \right)}\in \mathbb{C} ^{PQ\times1}$, defined element-wise by $\left[ \boldsymbol{h}_{spa}^{\left( i \right)} \right] _{pq}\triangleq \left[ \mathbf{P}_{pq} \right] _{m_i,n_i}$, which is given by:
\begin{equation}
    \boldsymbol{h}_{spa}^{\left( i \right)}=\alpha _i\left( \boldsymbol{a}_{\mathrm{Tx}}\left( \theta _i,\varphi _i \right) ^H\boldsymbol{w} \right) \boldsymbol{a}_{\mathrm{Rx}}\left( \theta _i,\varphi _i \right) +\boldsymbol{n}_i,
\end{equation}
where $\boldsymbol{n}_i$ denotes the additive noise vector.

The angle estimation problem can be formulated as a sparse recovery task. The sparse signal model is given by
\begin{equation}
    \boldsymbol{h}_{spa}^{\left( i \right)}=\mathbf{B}\boldsymbol{x}_i+\boldsymbol{n}_i,
\end{equation}
where $\mathbf{B}=\left[ \boldsymbol{b}\left( \theta _1,\varphi _1 \right) ,\boldsymbol{b}\left( \theta _2,\varphi _2 \right) ,...,\boldsymbol{b}\left( \theta _G,\varphi _G \right) \right] \in \mathbb{C} ^{PQ\times G}$ is an overcomplete dictionary whose columns are the effective steering vectors $\boldsymbol{b}\left( \theta ,\varphi \right) =\left( \boldsymbol{a}_{\mathrm{Tx}}\left( \theta,\varphi\right) ^H\boldsymbol{w} \right) \boldsymbol{a}_{\mathrm{Rx}}\left( \theta,\varphi \right) $ sampled on a discretized angular grid and $\boldsymbol{x}_i\in \mathbb{C} ^{G\times 1}$ is a sparse vector. 

To address this task, we propose the Zoom-OMP algorithm, which combines the idea of hierarchical processing with the OMP algorithm. The algorithm is an iterative process. Let $\boldsymbol{r}_0=\boldsymbol{h}_{spa}^{\left( i \right)}$ be the initial residual, $\Lambda _0=\emptyset $ be the initial support set, and $\mathbf{\Psi }_0=[] $ be the initial atom matrix. In the $t$-th iteration, the following steps are executed:

\textit{1) Coarse Search:} We first construct a low-resolution coarse dictionary $\mathbf{B}_{coarse}$ composed of effective steering vectors sampled with a large step size throughout the angular space. Then, we solve the following optimization problem to find the atom that is most correlated with the current residual $\boldsymbol{r}_{t-1}$, thus obtaining a coarse angle estimate:
\begin{equation}
    \left\{ \hat{\theta}_c,\hat{\varphi}_c \right\} =\underset{\left( \theta ,\varphi \right) \in Grid_{coarse}}{arg\max}\frac{\left| \boldsymbol{b}\left( \theta ,\varphi \right) ^H\boldsymbol{r}_{t-1} \right|}{\left\| \boldsymbol{b}\left( \theta ,\varphi \right) \right\| _2}.
\end{equation}
    
\textit{2) Fine Search:} Next, we construct a high-resolution fine dictionary $\mathbf{B}_{fine}$ only within a small neighborhood around the coarse angle estimate $\left( \hat{\theta}_c,\hat{\varphi}_c \right) $. A second correlation search is then performed to locate the atom index $\lambda_t$ that best matches the residual within this fine dictionary:
\begin{equation}
    \lambda_t=\underset{k\in Grid_{fine}}{arg\max}\frac{\left| \boldsymbol{b}_{k}^{H}\boldsymbol{r}_{t-1} \right|}{\left\| \boldsymbol{b}_k \right\| _2}.
\end{equation}

\textit{3) Support Set and Residual Update:} The selected atom index is added to the support set: $\Lambda_t=\Lambda_{t-1}\cup \left\{\lambda_t \right\} $. The selected atom vector is used to form the new atom matrix: $\mathbf{\Psi }_t=\left[ \mathbf{\Psi }_{t-1},\boldsymbol{b}_{\lambda _t} \right] $. Subsequently, the new signal estimate $\boldsymbol{x}_t$ is computed by orthogonal projection, and the residual $\boldsymbol{r}_t$ is updated:
\begin{equation}
    \boldsymbol{x}_t=\left( \mathbf{\Psi }_{t}^{H}\mathbf{\Psi }_t \right) ^{-1}\mathbf{\Psi }_{t}^{H}\boldsymbol{h}_{spa}^{\left( i \right)},
\end{equation}
\begin{equation}
\boldsymbol{r}_t=\boldsymbol{h}_{spa}^{\left( i \right)}-\mathbf{\Psi }_t\boldsymbol{x}_t.
\end{equation}

After $N_{target}$ iterations, the final support set $\mathbf{\Lambda }_{N_{target}}$ contains the indices, which can be used to retrieve the estimated angles from the discretized angular grid.

\subsection{Multi-BS Result Fusion}
The detection capability of a single BS is limited, as it can be affected by occlusions from other objects in the environment. This section, therefore, describes how to fuse the local point-cloud data obtained from various BSs into a unified global 3D point cloud.

First, a global coordinate system is defined for the entire sensing area, while each BS is associated with a local coordinate system whose origin is located at the antenna array center. All 4D parameters estimated in the preceding stages are described in the local coordinate system of each BS. To achieve multi-BS fusion, a coordinate transformation is required to register all local point clouds into the global coordinate system. We assume that the position (translation vector $\boldsymbol{t}_j\in \mathbb{R} ^{3\times 1}$) and orientation (rotation matrix $\mathbf{R}_j\in \mathbb{R} ^{3\times 3}$) of each BS $j\left( j=1,...,J \right) $ in the global coordinate system are known. For an arbitrary scatterer position $\boldsymbol{p}_{i}^{loc}$ estimated in the local coordinate system of BS $j$, its corresponding global position $\boldsymbol{p}_{i}^{glo}$ is obtained through the following rigid body transformation:
\begin{equation}
\boldsymbol{p}_{i}^{glo}=\mathbf{R}_j\boldsymbol{p}_{i}^{loc}+\boldsymbol{t}_j.
\end{equation}

After applying this coordinate transformation to all points in the local point cloud sets $\mathrm{P}_{j}^{loc}$ generated by each BS, we aggregate them to obtain the final global point cloud $\mathrm{P}_{glo}$:
\begin{equation}
    \mathrm{P}_{glo}=\bigcup_{j=1}^J{\left\{ \mathbf{R}_j\boldsymbol{p}_{i}^{loc}+\boldsymbol{t}_j \middle| \forall \boldsymbol{p}_{i}^{loc}\in \mathrm{P}_{j}^{loc} \right\}}.
\end{equation}
The fused global point cloud alleviates the limitations of  single-viewpoint sensing. It features higher point density and stronger robustness, and constitutes the final output of the environment reconstruction and imaging framework.

\section{NUMERICAL AND SIMULATION RESULTS}
Based on the proposed imaging scheme for the ISAC system, this section first introduces the key system parameters of the 5G NR waveform, followed by a qualitative and quantitative analysis of the imaging results.

\subsection{Simulation Parameter Setting}
\begin{table}[htbp]
\centering
\caption{Simulation System Parameters}
\label{tab:sim_params}
\begin{tabular}{ll}
\toprule
\textbf{Parameter Name}           & \textbf{Value or Specification}        \\
\midrule
Carrier Frequency                 & 26 GHz                                 \\
Subcarrier Spacing                & 120 kHz                                \\
Number of Resource Blocks (RBs)   & 264                                    \\
Bandwidth                         & 380 MHz                                \\
Number of Subframes               & 2                                      \\
Number of OFDM Symbols            & 224                                    \\
Number of Base Stations(BSs)      & 4                                      \\
Transmit/Receiver Antenna Array   & 8×8 Uniform Planar Array (UPA)         \\ 
SNR                               & 10 dB                                  \\

\bottomrule
\end{tabular}
\end{table}

\subsection{Imaging Performance}
\begin{figure}[t!]
    \centering 

    \begin{subfigure}[b]{0.49\columnwidth}
        \centering
        \includegraphics[width=\linewidth]{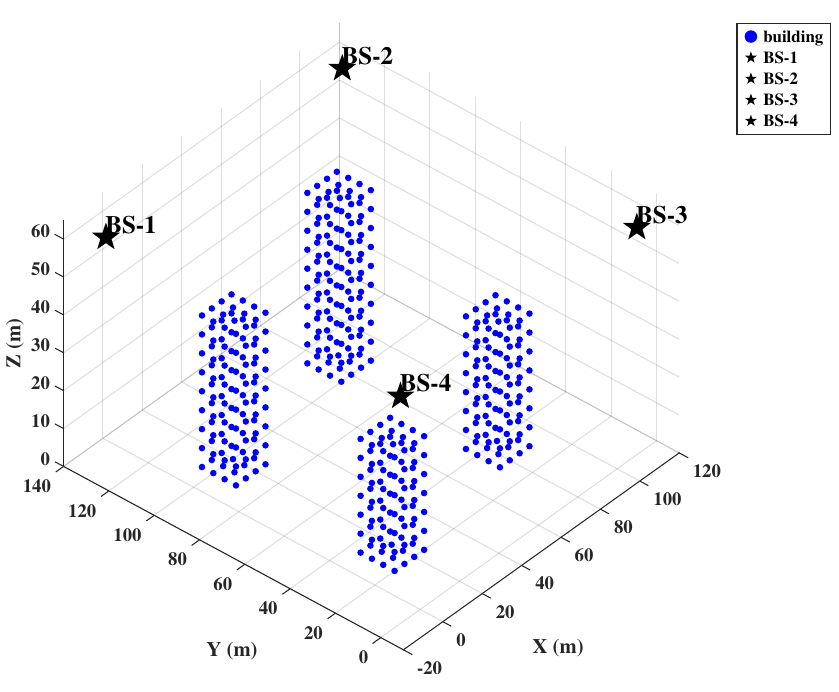} 
        \subcaption{Urban building scene (Ground Truth).}
        \label{fig:building_gt}
    \end{subfigure}
    \hfill 
    \begin{subfigure}[b]{0.49\columnwidth}
        \centering
        \includegraphics[width=\linewidth]{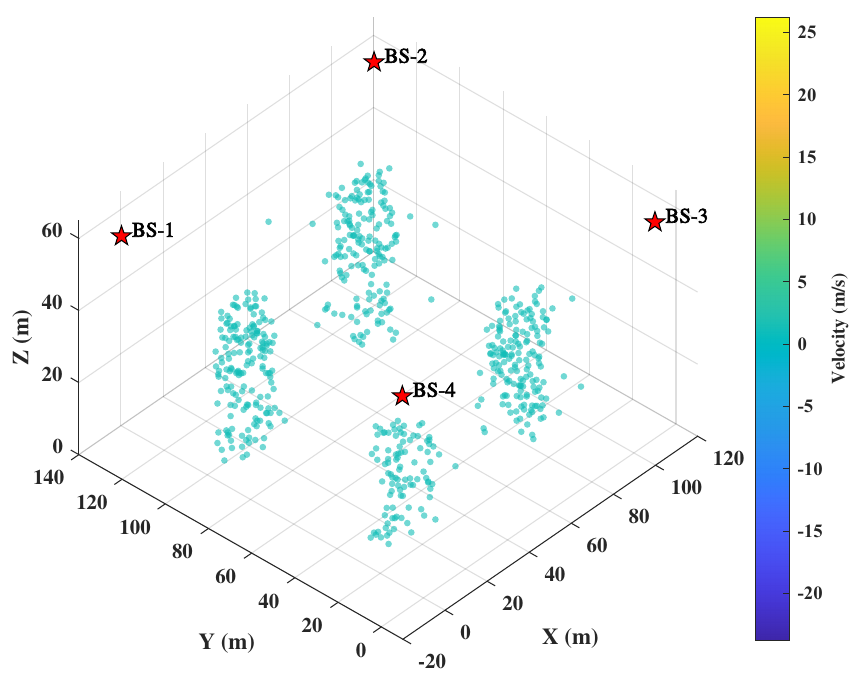} 
        \subcaption{Imaging result for the building scene.}
        \label{fig:building_result}
    \end{subfigure}
    
    \vspace{1em} 

    \begin{subfigure}[b]{0.49\columnwidth}
        \centering
        \includegraphics[width=\linewidth]{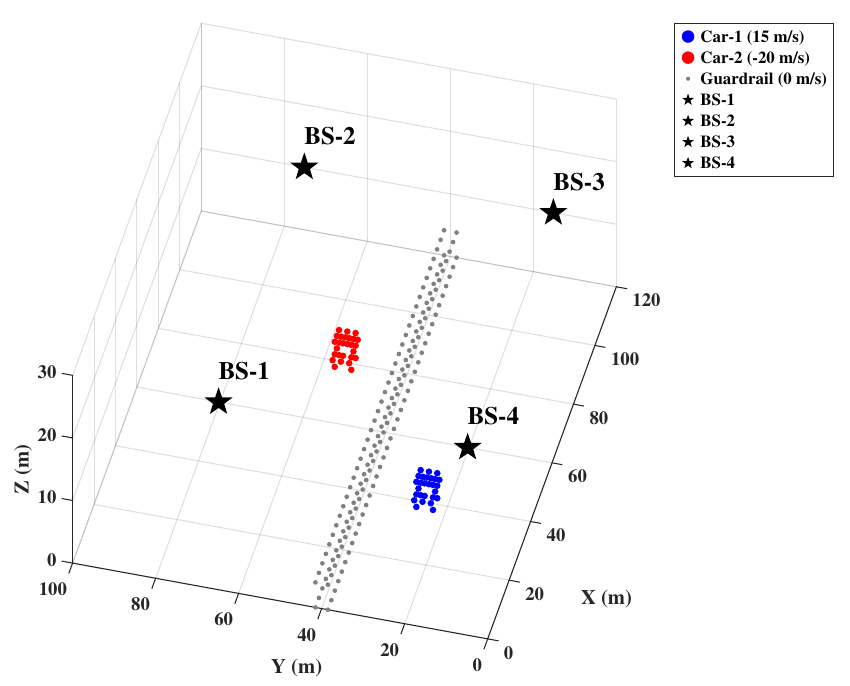} 
        \subcaption{Traffic road scene (Ground Truth).}
        \label{fig:road_gt}
    \end{subfigure}
    \hfill 
    \begin{subfigure}[b]{0.49\columnwidth}
        \centering
        \includegraphics[width=\linewidth]{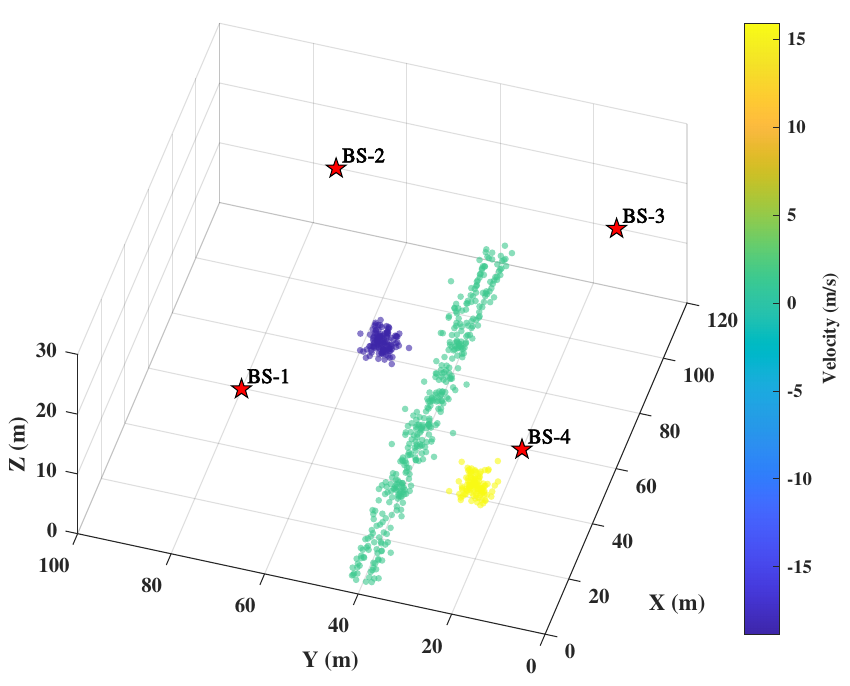} 
        \subcaption{Imaging result for the road scene.}
        \label{fig:road_result}
    \end{subfigure}

    \caption{Qualitative analysis of imaging results in two typical scenarios. Panels (a) and (c) show the ground truth models, while panels (b) and (d) present the corresponding point clouds generated by the proposed framework.}
    \label{fig:qualitative_results}
\end{figure}

The qualitative performance of the proposed imaging framework is demonstrated in Fig. \ref{fig:qualitative_results} for two representative scenarios. The strong correspondence between the generated point clouds (Fig. \ref{fig:building_result} and \ref{fig:road_result}) and their respective ground truth models (Fig. \ref{fig:building_gt} and \ref{fig:road_gt}) validates the proposed framework's capability to reconstruct both the 3D spatial structure and the velocity of targets in static and dynamic scenarios.

\subsection{Performance Comparison and Analysis}
To quantitatively evaluate the proposed framework, we employ the Chamfer Distance (CD) and F-score metrics.

The Chamfer Distance measures the average closest point distance between the ground truth point cloud $S_{\text{gt}}$, and the predicted point cloud $S_{\text{pred}}$. It is defined as:
\begin{align}
d_{\text{CD}}(S_{\text{gt}}, S_{\text{pred}}) = \frac{1}{|S_{\text{gt}}|} \sum_{x \in S_{\text{gt}}} \min_{y \in S_{\text{pred}}} \|x - y\|_2 + \\ \frac{1}{|S_{\text{pred}}|} \sum_{y \in S_{\text{pred}}} \min_{x \in S_{\text{gt}}} \|y - x\|_2 \notag,
\end{align}
where a lower CD value indicates a higher spatial accuracy of the reconstructed point cloud.

The F-score provides a balanced measure of imaging quality by considering both accuracy (Precision, $P$) and completeness (Recall, $R$). Precision represents the fraction of predicted points that are correctly located, while Recall represents the fraction of ground truth points that are successfully detected. The F-score is then calculated as :
\begin{equation}
\text{F-score} = 2 \cdot \frac{P \cdot R}{P + R},
\end{equation}
where a higher F-score (up to 1) signifies a better overall imaging quality.

\begin{figure}[t!]
    \centering
    
    \begin{subfigure}[b]{0.49\columnwidth}
        \centering
        \includegraphics[width=\linewidth, height=5cm, keepaspectratio]{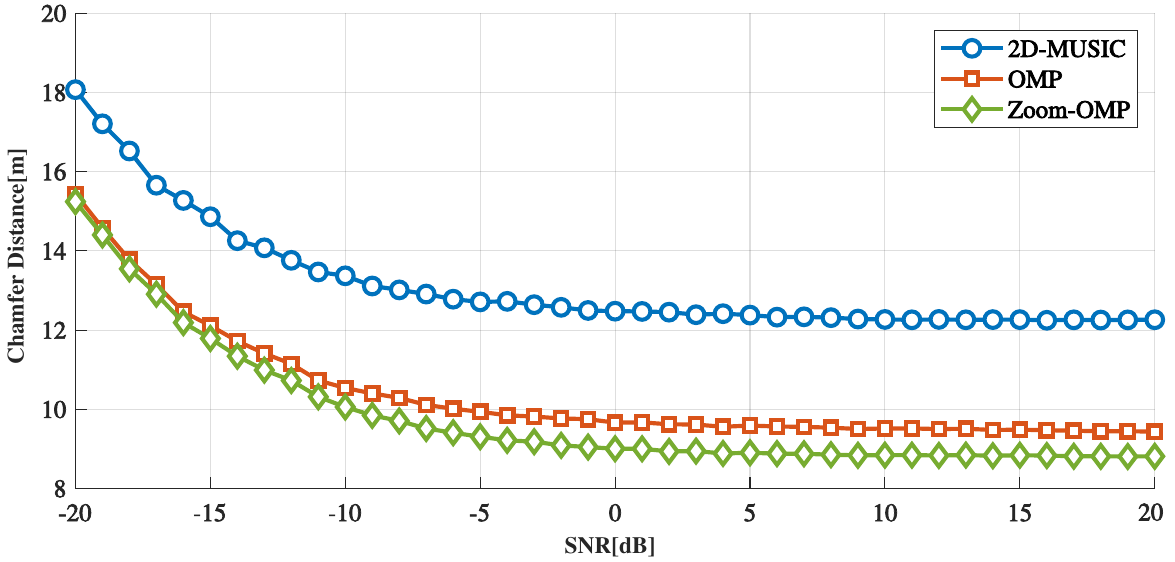} 
        \subcaption{Chamfer Distance vs. SNR.}
        \label{fig:chamfer_distance}
    \end{subfigure}
    \hfill
    \begin{subfigure}[b]{0.49\columnwidth}
        \centering
        \includegraphics[width=\linewidth, height=5cm, keepaspectratio]{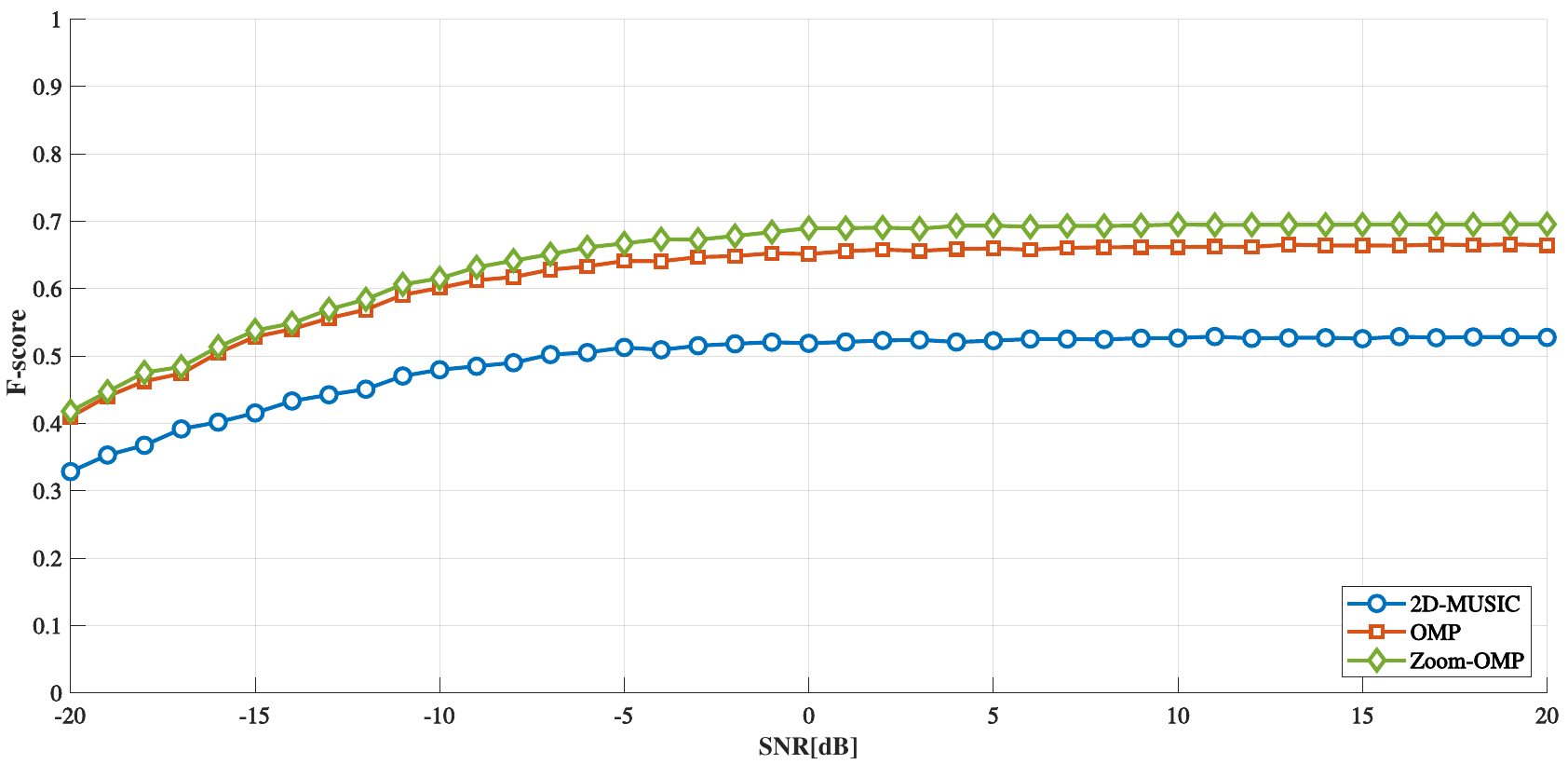} 
        \subcaption{F-score vs. SNR.}
        \label{fig:f_score}
    \end{subfigure}
    \caption{Quantitative performance comparison of different algorithms under varying SNR conditions.}
    \label{fig:quantitative_results}
\end{figure}

The quantitative results in Fig. \ref{fig:quantitative_results} show that our proposed Zoom-OMP algorithm achieves the best performance compared to the standard OMP and conventional 2D-MUSIC benchmarks. It consistently yields the lowest Chamfer Distance and the highest F-score across the entire SNR range, which validates its superior accuracy and robustness.

\subsection{Computational Complexity Analysis}

The computational complexity of Zoom-OMP is analyzed as follows. The standard OMP with a dictionary size $G$ requires $\mathcal{O}(PQG)$ operations per iteration. The 2D-MUSIC algorithm involves eigenvalue decomposition with $\mathcal{O}((PQ)^3)$ complexity. Zoom-OMP performs a coarse search over $G_c$ points and a fine search over $G_f$ points, where $G_c \ll G$ and $G_f \ll G$, resulting in the complexity $\mathcal{O}(PQ(G_c + G_f))$.

\section{CONCLUSION}
In this paper, we proposed a cooperative 4D imaging framework based on 5G NR frame structure. The framework operates in a two-stage process: First, high-resolution parameter estimation is performed at individual base stations, followed by a multi-BS data fusion stage. For the parameter estimation, we developed the Zoom-OMP algorithm, a computationally efficient method based on a coarse-to-fine sparse recovery strategy. Simulation results validated that the proposed framework achieves superior imaging accuracy and robustness compared to conventional benchmarks.

\bibliographystyle{IEEEtran}
\bibliography{references}

@article{liu2022integrated,
  title={Integrated sensing and communications: Toward dual-functional wireless networks for 6G and beyond},
  author={Liu, Fan and Cui, Yuanhao and Masouros, Christos and Xu, Jie and Han, Tony Xiao and Eldar, Yonina C and Buzzi, Stefano},
  journal={IEEE journal on selected areas in communications},
  volume={40},
  number={6},
  pages={1728--1767},
  year={2022},
  publisher={IEEE}
}

@article{tang2025cooperative,
  title={Cooperative ISAC-empowered low-altitude economy},
  author={Tang, Jun and Yu, Yiming and Pan, Cunhua and Ren, Hong and Wang, Dongming and Wang, Jiangzhou and You, Xiaohu},
  journal={IEEE Transactions on Wireless Communications},
  year={2025},
  publisher={IEEE}
}

@article{zhang2024target,
  title={Target localization in cooperative ISAC systems: A scheme based on 5G NR OFDM signals},
  author={Zhang, Zhenkun and Ren, Hong and Pan, Cunhua and Hong, Sheng and Wang, Dongming and Wang, Jiangzhou and You, Xiaohu},
  journal={IEEE Transactions on Communications},
  year={2024},
  publisher={IEEE}
}

@inproceedings{tan2021integrated,
  title={Integrated sensing and communication in 6G: Motivations, use cases, requirements, challenges and future directions},
  author={Tan, Danny Kai Pin and He, Jia and Li, Yanchun and Bayesteh, Alireza and Chen, Yan and Zhu, Peiying and Tong, Wen},
  booktitle={2021 1st IEEE International Online Symposium on Joint Communications \& Sensing (JC\&S)},
  pages={1--6},
  year={2021},
  organization={IEEE}
}

@article{guan20213,
  title={3-D imaging using millimeter-wave 5G signal reflections},
  author={Guan, Junfeng and Paidimarri, Arun and Valdes-Garcia, Alberto and Sadhu, Bodhisatwa},
  journal={IEEE Transactions on Microwave Theory and Techniques},
  volume={69},
  number={6},
  pages={2936--2948},
  year={2021},
  publisher={IEEE}
}

@inproceedings{lu2023isac,
  title={ISAC 4D imaging system based on 5G downlink millimeter wave signal},
  author={Lu, Bohao and Wei, Zhiqing and Wang, Lin and Zhang, Ruiyun and Feng, Zhingyong},
  booktitle={2023 IEEE Globecom Workshops (GC Wkshps)},
  pages={389--394},
  year={2023},
  organization={IEEE}
}

@inproceedings{zhou20236g,
  title={6G integrated sensing and communication-sensing assisted environmental reconstruction and communication},
  author={Zhou, Zhi and Li, Xianjin and He, Jia and Bi, Xiaoyan and Chen, Yan and Wang, Guangjian and Zhu, Peiying},
  booktitle={ICASSP 2023-2023 IEEE International Conference on Acoustics, Speech and Signal Processing (ICASSP)},
  pages={1--5},
  year={2023},
  organization={IEEE}
}

@inproceedings{song20243d,
  title={3D Environment Reconstruction Based on ISAC Channels},
  author={Song, Junzhe and He, Ruisi and Zhang, Zhengyu and Yang, Mi and Ai, Bo and Zhang, Haoxiang and Chen, Ruifeng},
  booktitle={2024 International Conference on Ubiquitous Communication (Ucom)},
  pages={487--491},
  year={2024},
  organization={IEEE}
}

@techreport{3gpp_ts38211_v17_5_0,
  author      = {{3GPP}},
  title       = {{5G; NR; physical channels and modulation}},
  institution = {{3rd Generation Partnership Project (3GPP)}},
  type        = {Technical Specification (TS)},
  number      = {38.211},
  month       = jul,
  year        = {2023},
  note        = {version 17.5.0.},
  url         = {https://www.etsi.org/deliver/etsi_ts/138200_138299/138211/}
}

@techreport{3gpp_ts38101_1_v17_10_0,
  author      = {{3GPP}},
  title       = {{5G; NR; user equipment (UE) radio transmission and reception; part 1: Range 1 stand alone}},
  institution = {{3rd Generation Partnership Project (3GPP)}},
  type        = {Technical Specification (TS)},
  number      = {38.101-1},
  month       = jul,
  year        = {2023},
  note        = {version 17.10.0.},
  url         = {https://www.etsi.org/deliver/etsi_ts/138100_138199/13810101/}
}

@techreport{3gpp_ts38101_2_v17_11_0,
  author      = {{3GPP}},
  title       = {{5G; NR; user equipment (UE) radio transmission and reception; part 2: Range 2 stand alone}},
  institution = {{3rd Generation Partnership Project (3GPP)}},
  type        = {Technical Specification (TS)},
  number      = {38.101-2},
  month       = oct,
  year        = {2023},
  note        = {version 17.11.0.},
  url         = {https://www.etsi.org/deliver/etsi_ts/138100_138199/13810102/}
}
 
\end{document}